\documentclass[12pt]{iopart}
\usepackage{iopams}
\usepackage{graphicx}
\begin{document}

\title[Ising model on hyperbolic lattice]{Ising model on hyperbolic lattice
studied by corner transfer matrix renormalization group method}

\author{R. Krcmar$^1$, A. Gendiar$^1$, K. Ueda$^2$ and T. Nishino$^2$}

\address{$^1$ Institute of Electrical Engineering, Centre of Excellence CENG,
Slovak Academy\\ of Sciences, D\'{u}bravsk\'{a} cesta 9, SK-841~04, Bratislava, Slovakia}
\address{$^2$ Department of Physics, Graduate School of Science, Kobe University,
Kobe 657-8501, Japan}
\ead{andrej.gendiar@savba.sk}

\begin{abstract}
We study two-dimensional ferromagnetic Ising model on 
a series of regular lattices, 
which are represented as a tessellation of polygons with $p \ge 5$ sides,
such as pentagons ($p = 5$), hexagons ($p = 6$), etc. 
Such lattices are on hyperbolic planes, which have constant negative
scalar curvatures. We calculate critical temperatures and scaling exponents 
by use of the corner transfer matrix renormalization group method. 
As a result, the mean-field like phase transition is observed for all the cases
$p \ge 5$. Convergence of the calculated transition 
temperatures with respect to $p$ is investigated towards the limit 
$p \rightarrow \infty$, where the system coincides with the Ising model on the 
Bethe lattice.
\end{abstract}

\submitto{\JPA}

\section{Introduction}
The Ising model has been extensively investigated because of its simplicity
in definition and wide
applicability to real magnetic materials. The model is exactly solvable
in two dimensions (2D) under appropriate conditions~\cite{Onsager,Baxter}.
For the study of insolvable cases, such as the cross-bond Ising model
and three-dimensional (3D) models, a variety of numerical methods have been
developed, such as Monte Carlo simulations~\cite{Binder}, Lanczos
diagonalization of row-to-row transfer matrices,
and Baxter's method of corner transfer matrices (CTMs)~\cite{Baxter}.
One of the recent technical progress in numerical study is establishment
of the density matrix renormalization group (DMRG)
method~\cite{White1,White2,Schollwoeck}.
The method is applicable to 2D classical lattice models including the Ising
model~\cite{Nishino1} and is of use for the study of higher-dimensional
lattice models~\cite{TPVA1,TPVA2,TT1,TT2,PEPS1,PEPS2}.

It is widely believed that the phase transition of the Ising model
belongs to the so-called Ising universality class provided that the system 
is uniform and on planar 2D lattices.
This universality can be violated if the lattice is in curved spaces, 
where typical examples are the lattices represented as regular 
tessellation of polygons in the hyperbolic plane, which has a 
constant negative scalar curvature~\cite{Rietman,Sausset,Doyon}.
As was pointed by Chris Wu {\it et al.}, boundary effects are non-negligible
below the transition temperature on
such hyperbolic lattices even in the thermodynamic limit~\cite{Chris1,Chris2}.
d'Auriac {\it et al.} investigated the bulk and boundary states
and discussed their difference~\cite{dAuriac}.
A recent Monte Carlo (MC) study by Shima
and Sakaniwa for the Ising model on one of the hyperbolic lattices shows that the critical
behavior in the ferromagnetic-paramagnetic transition deep inside the system
is mean-field like~\cite{Shima,Hasegawa}. Their result is in accordance with
the bulk property discussed by d'Auriac {\it et al.}~\cite{dAuriac}.

The size of the system treated by the MC simulations on the hyperbolic
lattices is limited by an exponential grow of the number of lattice points.
Some sort of renormalization group scheme is required under such a
situation. Quite recently we have applied the corner transfer matrix renormalization
group (CTMRG) method~\cite{Nishino2,Nishino3} to a particular hyperbolic
lattice which consists of pentagons ($p = 5$)~\cite{Ueda}. The
CTMRG method enables precise estimation of the bond energy and the
magnetization at the center of a sufficiently large system.
Ferromagnetic boundary conditions is assumed to observe the bulk property.
As a result, we have confirmed the mean-field like behavior of the
phase transition for the studied case $p = 5$.
In this article we extend our previous study by considering
hyperbolic lattices that consist of arbitrary ``$p$-gons''
with $p > 5$, such as hexagons ($p = 6$), heptagons ($p = 7$), etc. For the study
of large $p$ cases, we introduce a novel partial sum technique to the CTMRG method.

We calculate transition temperature $T_{\rm c}$ for each case 
$p \ge 5$ as well as related 
critical exponents $\alpha$, $\beta$ and $\delta$, respectively, associated with the specific 
heat, the spontaneous and induced magnetization. We then observe convergence 
of  $T_{\rm c}$ toward the limit $p \rightarrow \infty$, where the system corresponds to
the Ising model on the Bethe lattice.
In the next section we explain detail of the model 
on the hyperbolic lattices. We observe the structure 
of the lattices from the view point of the corner transfer matrix formalism. 
Numerical results are presented in Sec.~3, where we calculate the critical 
temperatures and the critical exponents. 
The conclusions are summarized in the last section.

\section{Structure of the system on hyperbolic lattice}

Consider a series of infinite-size lattices that consist of regular polygons with
$p \ge 5$ sides, which are called as `$p$-gons'. Each lattice is represented as
a tessellation of the $p$-gons on an infinite
plane with a constant negative scalar curvature. One can classify this
type of lattices by a pair of integers $(p,q)$, where the coordination number $q$
represents the number of the neighboring lattice points. In the following we
consider the $( p \! \ge \! 5, \, q \! = \! 4 )$ lattices, including
the pentagonal lattice $(5,4)$, the hexagonal one $(6,4)$,
the heptagonal one $(7,4)$, etc.
We also treat a square lattice $(4,4)$ defined  on the flat plane for
comparison. 

\begin{figure}
\centerline{\includegraphics[width=6.2cm,clip]{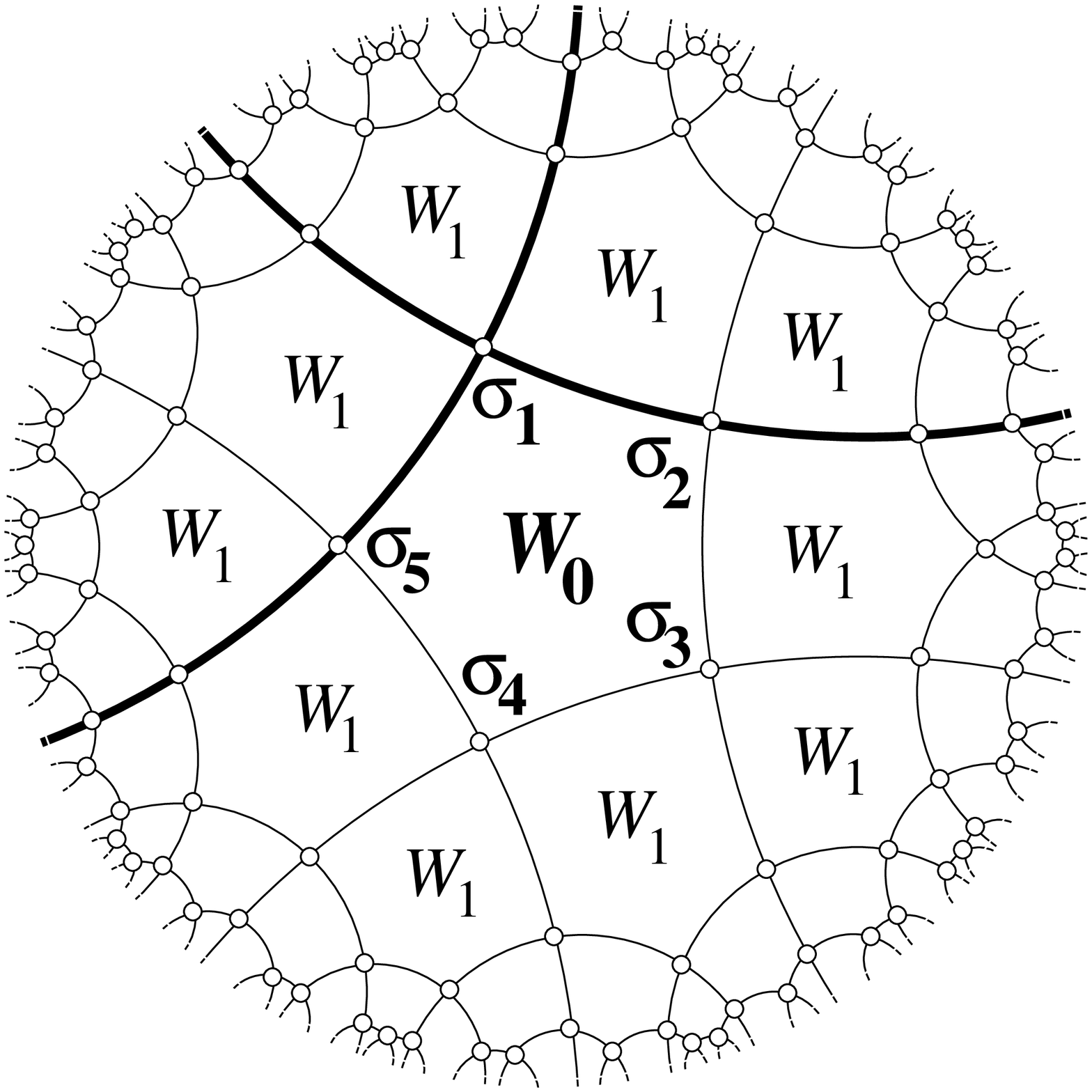}
\includegraphics[width=6.5cm,clip]{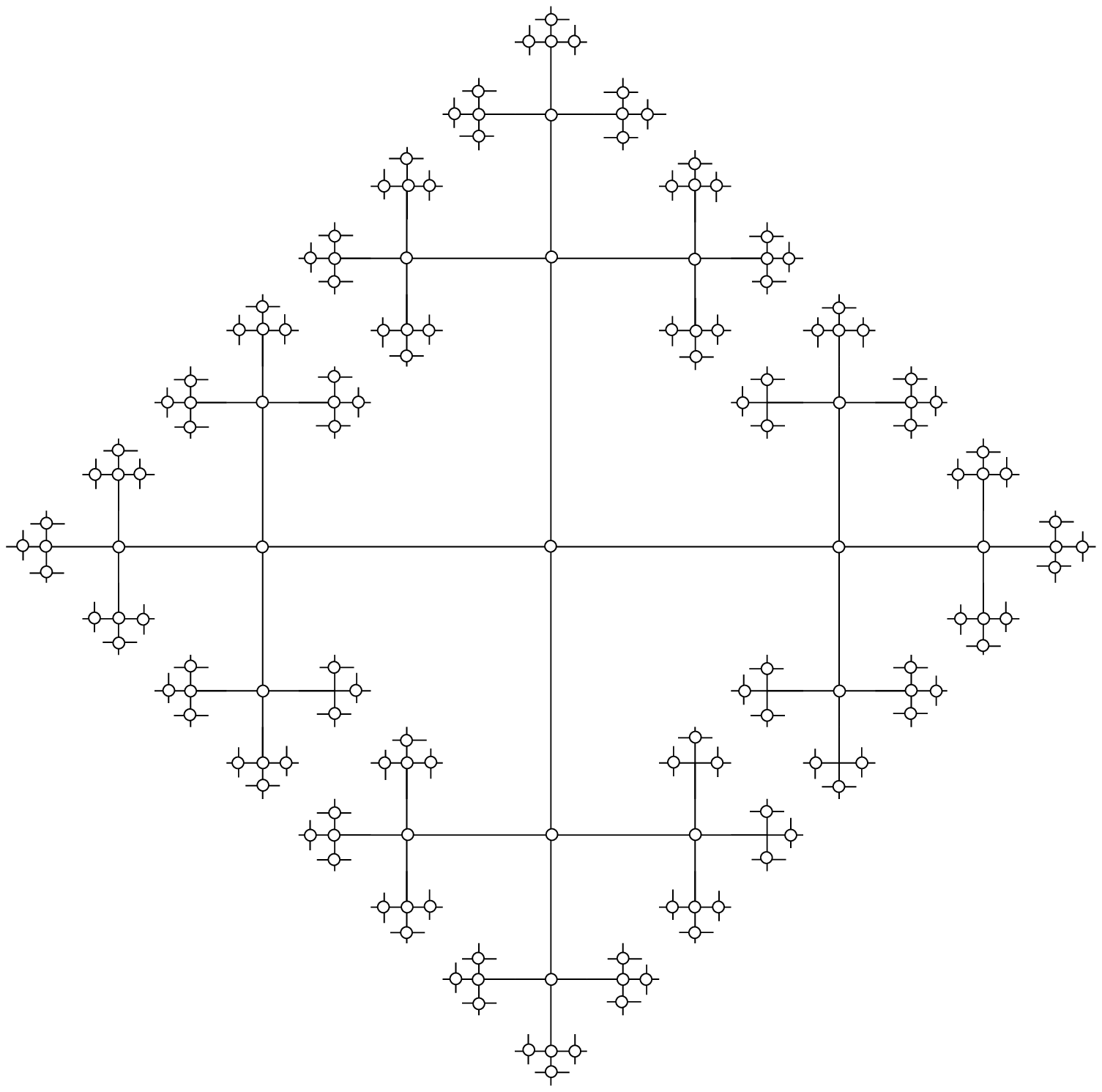}}
\caption{
Left: the Ising model on the pentagonal lattice $(5,4)$ which is
drawn in the Poincar\'{e} disc. The open circles represent the Ising
spins $\sigma_i^{~}$. Note that each pentagon has the same size
and shape. Right: the Bethe lattice of the coordination number $q=4$
is equivalent to the $(\infty,4)$ lattice.
}
\label{f1}
\end{figure}

As an example, we draw the pentagonal lattice $(5,4)$ in the left part of
Fig.~\ref{f1}, where the infinite area of the hyperbolic plane is mapped
into the Poincar\'{e} disc. All arcs in the
figure represent geodesics that are perpendicular to the bounding circle.
Two geodesics drawn by the thick arcs cross one another at a lattice point.
Note that by these two geodesics, the whole system is divided 
into four equivalent semi-infinite parts, which are called as the {\it quadrants} 
or {\it corners}. As another typical example, we draw the $( \infty, 4 )$ 
lattice in the right part of Fig.~\ref{f1}. This lattice is merely the Bethe lattice
with the coordination number $p = 4$. Note that the 
%
%
%
Hausdorff dimension of these 
 $( p \! \ge \! 5, \, 4)$ lattices is infinite.

Consider the Ising model on the $(p \! \ge \! 5, \, 4)$ lattice, where on each lattice point
there is an Ising spin $\sigma_i^{~} = \uparrow\downarrow$.
If only the neighboring Ising interactions are assumed, the Hamiltonian
of the system is represented as
\begin{equation}
{\cal H} = - J \sum_{\{i,j\}}\sigma_i^{~} \sigma_j^{~} - H\sum_{\{i\}}\sigma_i^{~} \, ,
\end{equation}
where the summation $\{i,j\}$ runs over all nearest-neighbor spin pairs.
We assume that the interaction is ferromagnetic ($J > 0$). The external
magnetic field $H$  acts on each spin site uniformly. 
For latter conveniences of expressing the partition function, let us introduce 
the weight $w(\sigma_i^{~} \, \sigma_j^{~} )$ assigned to the neighboring 
spin pair $\{i,j\}$ 
\begin{equation}
w(\sigma_i^{~} \, \sigma_j^{~} ) 
= \exp\left[ \beta J \frac{\sigma_i^{~} \sigma_j^{~}}{2}
+ \beta H \frac{\sigma_i^{~} + \sigma_j^{~}}{8}
\right]
\label{eq2_1}
\end{equation}
with $\beta = 1 / k_{\rm B}^{~} T$. The Boltzmann weight of the whole system
is then expressed as
\begin{equation}
\exp( - \beta {\cal H} ) = \prod_{ \{i,j\} }^{~} 
\left[ w(\sigma_i^{~} \, \sigma_j^{~} ) 
\right]^2_{~} \, .
\end{equation}
Since each bond is shared by two $p$-gons, it is possible to assign a
local Boltzmann weight for each $p$-gon. Let us focus on the $p$-gon,
where spins on its edges are labeled by
$\sigma_1^{~}$, $\sigma_2^{~}$, $\ldots$, and $\sigma_{p}^{~}$,
as shown in the left side of Fig.~\ref{f1}, where  the case $p = 5$ is drawn
as an example.
The Boltzmann weight assigned to the $p$-gon, which is called as the `face weight', 
is then expressed as
\begin{equation}
W(\sigma_1^{~} \, \sigma_2^{~} \, \sigma_3^{~} \, \dots \, \sigma_p^{~}) =
w(\sigma_1^{~} \, \sigma_2^{~}) \,
w(\sigma_2^{~} \, \sigma_3^{~}) \ldots
w(\sigma_{p-1}^{~} \, \sigma_p^{~}) \, 
w(\sigma_{p}^{~} \, \sigma_1^{~}) \, .
\label{eq2}
\end{equation}
It is straightforward that one can assign the same weight $W$ for all the
$p$-gons in the system. 
We have thus represented the Ising model on the $(p \! \ge \! 5, \, 4)$ 
lattices as a special case of the interaction-round-a-face (IRF) model,
which regards the `face' as the unit of the system~\cite{Baxter}.

The partition function of a finite-size system is represented as
\begin{equation}
{\cal Z} = \sum_{\{\sigma\}}\prod W \, ,
\end{equation}
where the sum is taken over all configurations of the spins.
The product runs over all the face weights contained in the system
starting from a weight, which is shown as $W_0^{~}$ on the left in Fig.~1,
at the center of the system. Around $W_0^{~}$ there are $2p$ number
of neighboring  weights $W_1^{~}$ in the first shell, $4p(p-3)$ number of
$W_2^{~}$ in the second shell, etc. The number
of the weights and sites in the $\alpha$-th shell increases exponentially with $\alpha$.

For the calculation of the partition function ${\cal Z}$, we introduce
the corner transfer matrix (CTM) denoted by $C$ that represents the 
Boltzmann weight for each quadrant of the system~\cite{Baxter}. By use of
the CTM, the partition function is expressed as the
trace
\begin{equation}
{\cal Z} = {\rm Tr}  \, C^4_{~} 
\end{equation}
of the density matrix $\rho = C^4_{~}$.
In the following we use the common notations in the CTMRG
method~\cite{Nishino2,Nishino3,Ueda}; see the detail in Ref.~\cite{Ueda}.

\begin{figure}
\centerline{\includegraphics[width=5.5cm,clip]{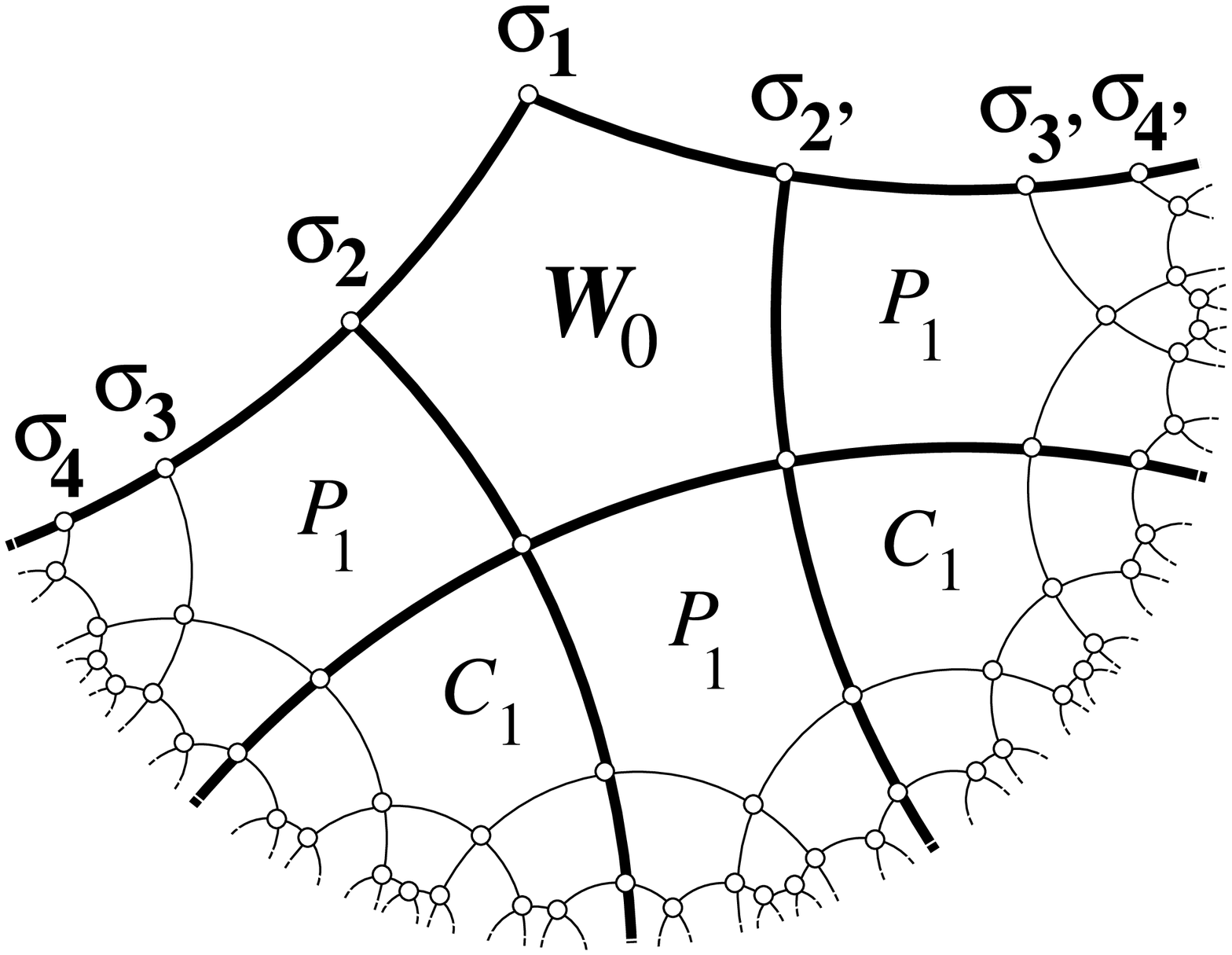}
\includegraphics[width=5.0cm,clip]{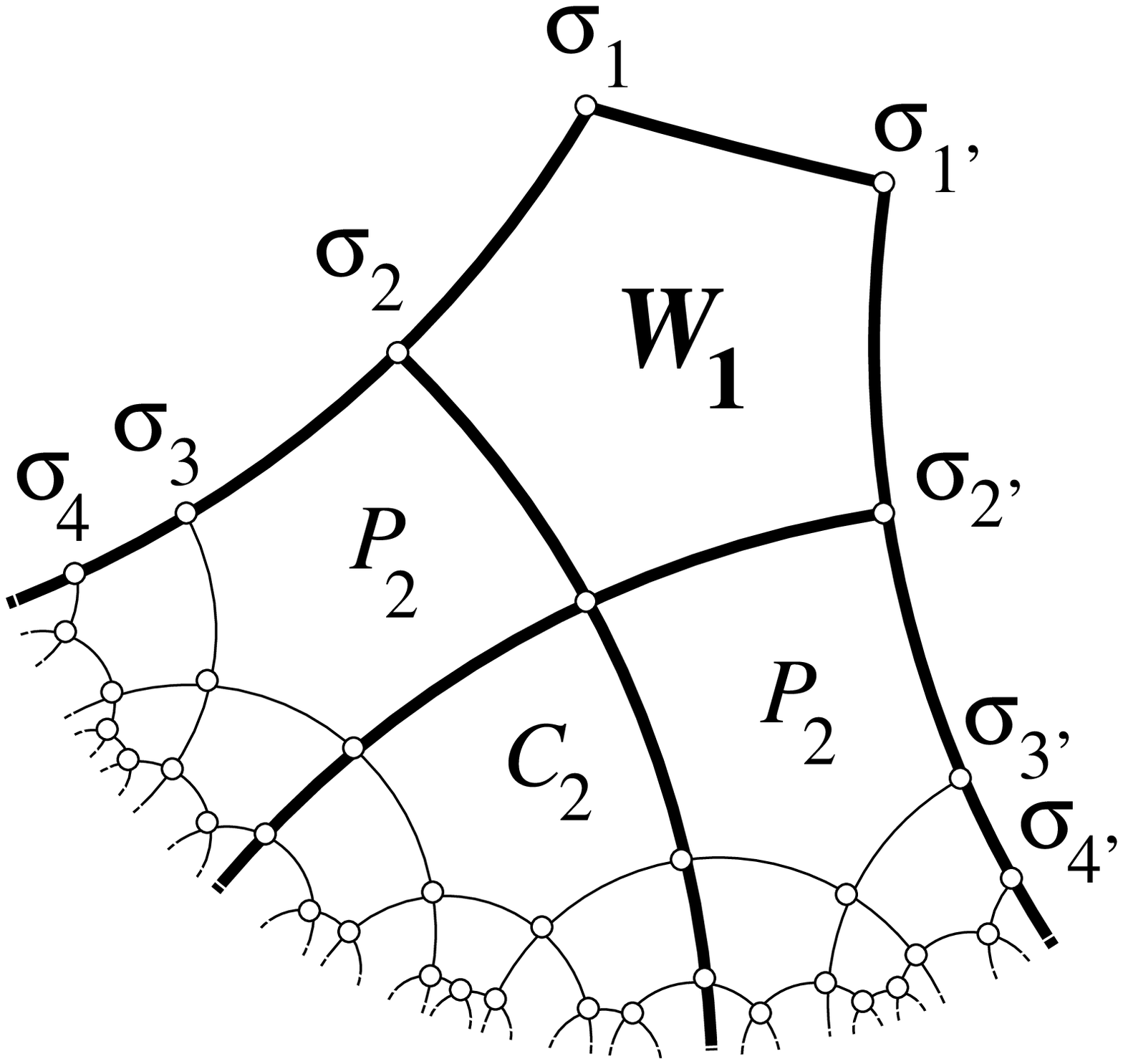}}
\caption{
A corner transfer matrix 
$C(\dots\sigma_3^{~}\sigma_2^{~}\sigma_1^{~} | 
\sigma_{1^\prime}^{~}\sigma_{2^\prime}^{~}
\sigma_{3^\prime}^{~}\dots)$ 
of the case $p = 5$ shown on the left side 
consists of a face weight $W_0^{~}$, two
CTMs of smaller size $C_1^{~}$, and three half-row transfer
matrices $P_1^{~}$. Each HRTM 
$P(\dots\sigma_3^{~}\sigma_2^{~}\sigma_1^{~} | 
\sigma_{1^\prime}^{~}
\sigma_{2^\prime}^{~}\sigma_{3^\prime}^{~}\dots)$ shown
in the right has an analogous substructure.
}
\label{f2}
\end{figure}

Let us consider a finite-size system that contains the lattice points up to the
$N$-th shell, where the ferromagnetic boundary
condition is imposed at the lattice border.
The left side of Fig.~\ref{f2} shows the structure of the CTM
of the system for the case $p = 5$. The CTM $C$ contains
a face weight labeled by $W_0^{~}$, two CTMs of the smaller size 
labeled by $C_1^{~}$, and
three parts labeled by $P_1^{~}$ that corresponds to the so-called 
half-raw transfer matrix (HRTM). The right side of Fig.~\ref{f2} shows
similar substructure of the HRTM for $p = 5$.
Looking at these figures, one finds a recursive relation between the
CTMs and the HRTMs. If one has $C$ and $P$ of a certain linear size, one can
obtain the extended ones $C'$ and $P'$ by the following fusion process~\cite{Ueda}
\begin{eqnarray}
C' &=& W \cdot P \cdot \left( C \cdot P \right)^{p-3}_{~} \,  \nonumber\\
P' &=& W \cdot P \cdot \left( C \cdot P \right)^{p-4}_{~} \, ,
\label{itereq}
\end{eqnarray}
which increases the linear size of $C$ and $P$ by one. 
Note that if ferromagnetic boundary condition is imposed for both 
 $C$ and $P$, the extended ones $C'$ and $P'$ are also subject to the
same boundary condition. Repeating this fusion
process, one can obtain CTMs and HRTMs of arbitrary linear sizes
provided that these matrices can be stored to a computational machine.
This storage limitation can be removed by use of the renormalization group (RG)
transformation in the density matrix scheme~\cite{White1,White2,Schollwoeck}. 
As a result, the matrices $C$ and $P$ are {\it renormalized}
into effective ones ${\tilde C}$ and ${\tilde P}$, whose
matrix dimension is at most $2m$ where $m$ is
the number of states kept for each block spin~\cite{White1}.

One-point functions at the center of the system are easily calculated 
by use of ${\tilde C}$ thus obtained by way of sufficient number of iterative 
extensions and the RG transformations.
For example, the spontaneous magnetization is calculated as
\begin{equation}
{\cal M} = \langle\sigma\rangle
= \frac{ {\Tr}~\sigma \, {\tilde C}_{~}^4}{{\Tr}~
{\tilde C}_{~}^4} \, ,
\end{equation}
where $\sigma$ denotes the Ising spin at the center of the
system. For the bond energy, we similarly express it as
\begin{equation}
\label{ie}
{\cal U} = - J \langle\sigma\tau\rangle
= - J \, \frac{{\Tr}~\sigma\tau \,  {\tilde C}_{~}^4
}{{\Tr}~{\tilde C}_{~}^4} \, ,
\end{equation}
where $\tau$ is a neighboring spin to $\sigma$.
From the calculated ${\cal U}$, the specific heat can be obtained by taking
the numerical differential ${\cal C}=\partial {\cal U}/\partial T$.

\section{Numerical Results}

Numerical analysis is carried out for the cases $p \ge 5$. Because of the
product structure of the local weight $W$ shown in Eq.~\ref{eq2}, the
fusion process expressed by Eq.~\ref{itereq} can be performed
for arbitrary large $p$ without any increase of computational memory.
We keep at most $m = 50$ states for the 
block spin variable during the CTMRG calculations.
For all the cases investigated here, the density matrix eigenvalues
decay very fast even at the transition temperature. 
This is in contrast to the relatively slow decay observed in the square
lattice models~\cite{Okunishi}. Thus actually $m = 10$
is sufficient for the calculation of the magnetization
${\cal M}$ as well as the bond energy ${\cal U}$.
\begin{figure}
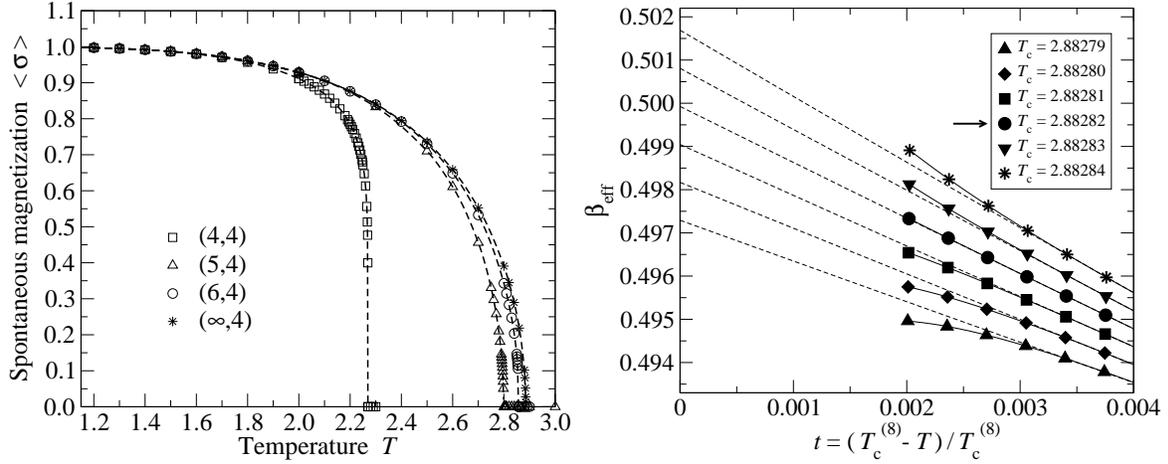

\centerline{\includegraphics[width=7.5cm,clip]{Figure3a.eps}
\includegraphics[width=7.7cm,clip]{Figure3b.eps}}
\caption{
Left: the spontaneous magnetization ${\cal M}$
with respect to temperature $T$ at $H=0$. Right: the $t$-dependence
of the effective critical exponent in Eq.~(\ref{efex}) for the case
of $(8,4)$ lattice.
}
\label{f3}
\end{figure}

The left side of Fig.~\ref{f3} shows the temperature dependence of 
the spontaneous magnetization.
We have chosen dimensionless parameters
$k_{\rm B} = J = 1$. For comparison, we also draw ${\cal M}$ for the
case of the Bethe lattice with the coordination number $q=4$.
In the critical region below the transition temperature $T_{\rm c}^{(p)}$,
the magnetization behaves as ${\cal M} = f( t ) \, t^\beta_{~}$,
where $f( t )$ is a slowly varying function of $t = (T_{\rm c}^{(p)}-T)/T_{\rm c}^{(p)}$, 
the rescaled temperature deviation from $T_{\rm c}^{(p)}$.
In order to estimate $T_{\rm c}^{(p)}$ precisely,  we
plot the effective critical exponent
\begin{equation}
\beta_{\rm eff}( t ) = \frac{\partial}{\partial\ln t} \ln{\cal M} 
= \beta + \frac{\partial}{\partial\ln t} \ln \, f( e^{\ln t}_{~} ) 
= \beta + \frac{f'}{f} \, t + \ldots
\label{efex}
\end{equation}
in a very small $t$ region.
The right side of Fig.~\ref{f3} shows the effective
exponent $\beta_{\rm eff}(t)$ thus calculated  for the case $p = 8$.
From the trial critical temperatures listed in the inset, 
$T_{\rm c}^{(8)}=2.88282$ gives the best linear fit.
We have applied the same procedure for all $p$ that we have chosen.
The results are listed in Table~\ref{t1}, where
\begin{table}[t]
\caption{
\label{t1} The calculated critical temperatures $T_{\rm c}^{(p)}$.
}
\begin{center}
\begin{tabular}{lcccccc}
\br
$( p, q )$ & $(4,4)$ & $(5,4)$ & $(6,4)$ & $(7,4)$ & $(8,4)$ & $(9,4)$ \\
\mr
$T_{\rm c}^{(p)}$ & $2/\ln\left(\sqrt{2}+1\right)$ & 2.79908 & 2.86050 & 
2.87754 & 2.88282 & 2.88457 \\
\br
$( p, q )$ & $(10,4)$ & $(11,4)$ & $(12,4)$ & $(15,4)$ & $(30,4)$ & $(\infty,4)$ \\
\mr
$T_{\rm c}^{(p)}$ & 2.88519 & 2.88533 & 2.88538 & 2.88539 & 2.88539 & $1/\ln\sqrt{2}$ \\
\br\end{tabular}
\end{center}
\end{table}
\begin{figure}
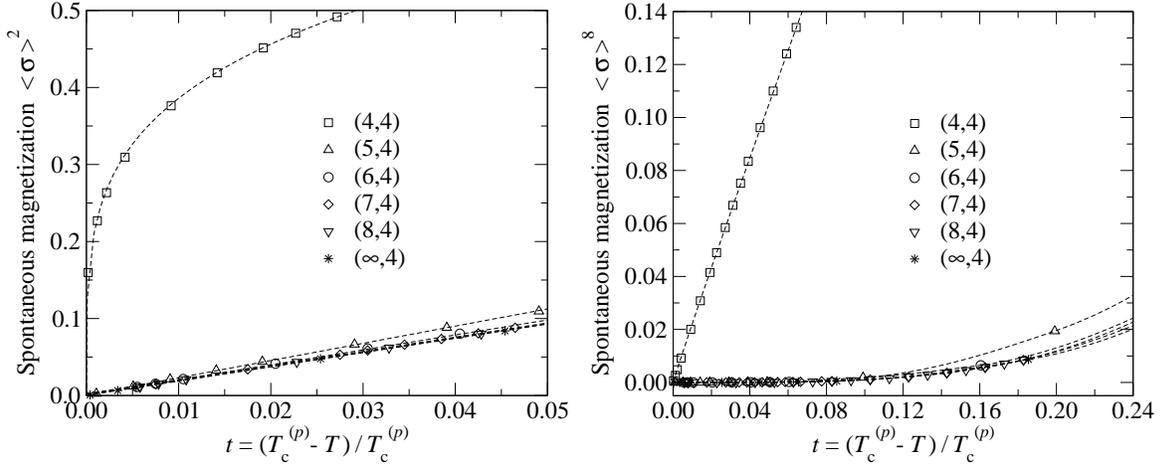

\centerline{
\includegraphics[width=7.5cm,clip]{Figure4a.eps}
\includegraphics[width=7.65cm,clip]{Figure4b.eps}}
\caption{
The $t$-dependence of ${\cal M}^2_{~}$ (left) and ${\cal M}^8_{~}$ (right).
The mean-field exponent $\beta = \frac{1}{2}$ is observed for $p \ge 5$,
whereas $\beta = \frac{1}{8}$ exclusively for $p = 4$.}
\label{f4}
\end{figure}
$\beta_{\rm eff}(0) = \beta\cong\frac{1}{2}$ is confirmed for all the cases.
Figure~\ref{f4} shows the $t$-dependence of ${\cal M}^2_{~}$ (left) and
${\cal M}^8_{~}$ (right). It is obvious that the mean-field exponent
$\beta = \frac{1}{2}$ is observed for all the cases $p\ge5$,
whereas the Ising universality class $\beta=\frac{1}{8}$ is realized
for the square lattice $(4,4)$ only.

\begin{figure}
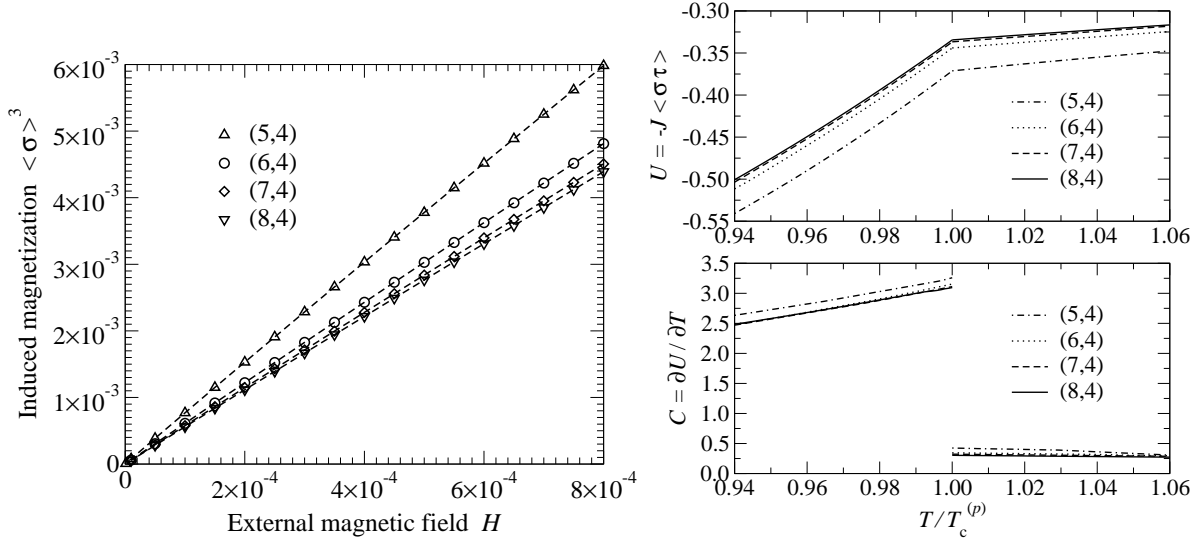

\centerline{\includegraphics[width=8.4cm,clip]{Figure5a.eps}
\includegraphics[width=7.2cm,clip]{Figure5b.eps}
}
\caption{
Left: Induced magnetization at $T_{\rm c}^{(p)}$ with respect to
the applied magnetic field $H$.
Right: the upper and lower panels, respectively, display singularity
of the internal energy ${\cal U}$ and the specific heat ${\cal C}$ around
in the critical region.}
\label{f5}
\end{figure}

At the calculated $T_{\rm c}^{(p)}$, let us observe the induced magnetization
${\cal M}$ with respect to the applied field $H$. From the
scaling relation ${\cal M} \propto H^{1/\delta}_{~}$, 
another critical exponent $\delta$ can be extracted. The left side of 
Fig.~\ref{f5} shows the linearity of ${\cal M}^{3}_{~}$
with respect to small external magnetic fields $H$ calculated at the 
critical temperature $T_{\rm c}^{(p)}$ listed in Table 1. It is apparent 
that $\delta$ is equal to $3$, which supports the mean-field like
behavior of the Ising model on the $(p,4) $ lattices when $p\ge5$. 

To confirm the mean-field nature of the phase transition, we calculate
the internal energy ${\cal U}$ by way of Eq.~(\ref{ie}).
The right side of Fig.~\ref{f5} shows  ${\cal U}$ with respect to the rescaled
temperature $T / T_{\rm c}^{(p)}$. For each case
there is a cusp at $T = T_{\rm c}^{(p)}$, and a linear dependence
of ${\cal U}$ in the vicinity of $T_{\rm c}^{(p)}$ supports
the critical exponent $\alpha=0$. There is a jump in specific heat.

\begin{figure}
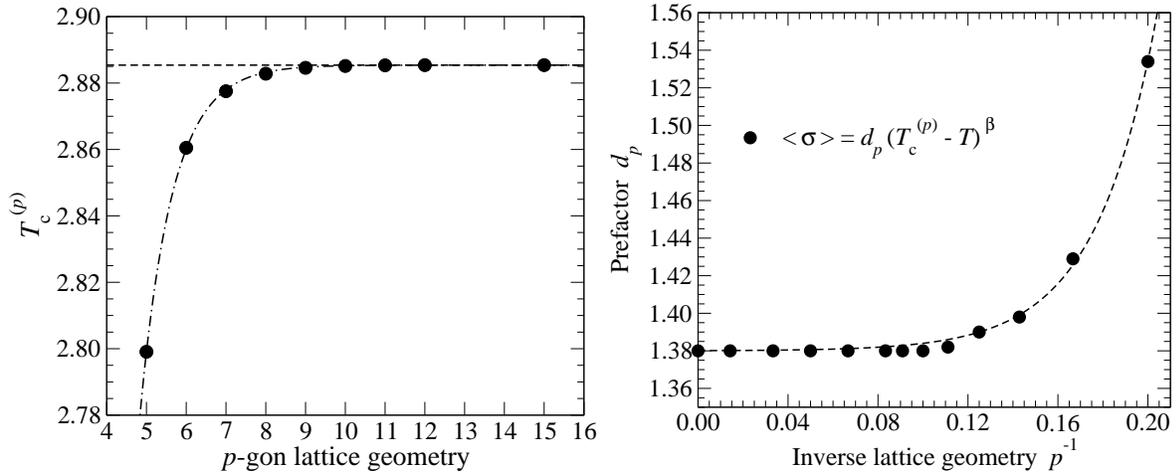

\centerline{
\includegraphics[width=7.85cm,clip]{Figure6a.eps}
\includegraphics[width=7.5cm,clip]{Figure6b.eps}}
\caption{
Left: the exponential dependence of $T_{\rm c}^{(p)}$
with respect to $p$. The dashed horizontal line corresponds to the exact
result on the Bethe lattice.
Right: a scaling law of the prefactor $d_p^{~}$ associated with
temperature dependence of the spontaneous magnetization with
respect to $p$.
}
\label{f6}
\end{figure}

Let us observe the convergence of $T_{\rm c}^{(p)}$ 
with respect to $p$ towards $T_{\rm c}^{(\infty)} = 1/\ln\sqrt{2} = 2.88539$. 
As shown on the left side of Fig.~\ref{f6}, 
the convergence is exponential 
\begin{equation}
T_{\rm c}^{(p)} - T_{\rm c}^{(\infty)} \propto  e^{- a p}_{~} \, 
\end{equation} 
with respect to $p$. Fitting the plotted data for $5 \le p \le 8$,
we have obtained the decay factor $a = 1.2543$. The 
prefactor $d_p^{~}$ in the scaling relations
\begin{equation}
{\cal M} = d_p^{~} \, ( T_{\rm c}^{(p)} - T )^{\beta}_{~} 
\end{equation}
also shows a monotonous convergence to $d_{\infty}^{~}$ as shown on
the right side of Fig.~\ref{f6}. We have not obtained any appropriate
fitting function of the $p$-dependence yet (the dashed line corresponds
to an exponential fit).

\section{Conclusions}

We have calculated the magnetization, the internal energy and
the specific heat of the Ising model on a series of 
$( p \! \ge \! 5, \, 4 )$ lattices on the
hyperbolic planes. These quantities are observed at the center of
the system with ferromagnetic boundary condition.
We calculated the critical exponents and
obtained $\alpha=0$, $\beta=\frac{1}{2}$, 
and $\delta=3$ for all the cases.
Our result supports and complements previous predictions given by 
d'Auriac {\it et al.}~\cite{dAuriac}, and independently by 
Shima {\it et al.}~\cite{Shima,Hasegawa}. The obtained
results are in accordance
with the fact that the Hausdorff dimension is infinite on the hyperbolic lattices
and also with common knowledge that the mean-field like phase transition is
observed above the critical dimension $d_{\rm c}=4$~\cite{Wu}.

The transition temperature $T_{\rm c}^{(p)}$ of the Ising model on the $( p, 4 )$
lattice converges exponentially fast towards $T_{\rm c}^{(\infty)}$
with respect to increasing $p$. We have not yet clarified physical interpretation
of this convergence. A renormalization group scheme given by Hilhorst {\it et al.}
may provide some  information to this question~\cite{Leewen}. A recent numerical 
renormalization group scheme suggested by Levin and Nave might be of use to find out an
appropriate fixed-point Hamiltonian~\cite{Levin}.

Recent study of the planar rotator (i.e. the classical XY) model on a
hyperbolic lattice suggests that the mean-field like phase transition is not
always realized for systems with the hyperbolic geometry~\cite{Baek}.
Such XY model can be investigated by the generalized CTMRG method
explained in this article~\cite{Ueda} if appropriate boundary conditions are
chosen~\cite{Takasaki}.

\ack
This work is partially supported by Slovak Agency for Science and Research 
grant APVV-51-003505 and Slovak VEGA grant No. 2/6101/27 (A.G. and R.K.) 
as well as partially by a Grant-in-Aid for Scientific Research from Japanese 
Ministry of Education, Culture, Sports, Science and Technology (T.N. and A.G.).

\end{document}